\documentclass[12pt]{iopart}
\usepackage{graphicx}
\usepackage{iopams}
\expandafter\let\csname equation*\endcsname\relax
\expandafter\let\csname endequation*\endcsname\relax
\usepackage{amssymb}
\usepackage{amsmath}
\usepackage{multirow}
\usepackage{xcolor}
\usepackage{array}
\usepackage{textcomp}
\newcommand{\ped}[1]{\ensuremath{_{\rm #1}}}

\begin{document}

\title{Turning non-superconducting elements into superconductors by quantum confinement and proximity}

\author{Giovanni Alberto Ummarino$^{1,2}$ and Alessio Zaccone$^{3,4}$}
\address{$^1$ Istituto di Ingegneria e Fisica dei Materiali,
Dipartimento di Scienza Applicata e Tecnologia, Politecnico di
Torino, Corso Duca degli Abruzzi 24, 10129 Torino, Italy. ORCID 0000-0002-6226-8518}
\address{$^2$ National Research Nuclear University MEPhI (Moscow Engineering
Physics Institute), Kashirskoe shosse 31, Moscow 15409, Russia}
\address{$^3$ Department of Physics ``A. Pontremoli'', University of Milan, via Celoria 16,
20133 Milan, Italy}
\address{$^4$ Institut f{\"u}r Theoretische Physik, University of G{\"o}ttingen,
Friedrich-Hund-Platz 1,
37077 G{\"o}ttingen, Germany}
\ead{giovanni.ummarino@polito.it}
\ead{alessio.zaccone@unimi.it}

\begin{abstract}
Elemental good metals, including noble metals (Cu, Ag, Au) and several $s$-block elements, do not exhibit superconductivity in bulk at ambient pressure, primarily due to weak electron--phonon coupling that fails to overcome Coulomb repulsion. Quantum confinement in ultra-thin films is known to strongly reshape the electronic spectrum and the density of states near the Fermi level, and in established superconductors it produces pronounced, often non-monotonic, thickness dependencies of the critical temperature.  
In this perspective, we examine whether quantum confinement alone, or in combination with proximity effects, can induce an observable superconducting instability in metals that are non-superconducting in bulk form. We review recent theoretical progress and present a unified framework based on a confinement-generalized, isotropic one-band Eliashberg theory, in which the normal density of states becomes energy dependent and key material parameters ($E_F$, $\lambda$, and $\mu^{*}$) acquire an explicit thickness dependence.  
By numerically solving the resulting Eliashberg equations using ab-initio or experimentally determined electron--phonon spectral functions $\alpha^{2}F(\Omega)$ and Coulomb pseudopotentials $\mu^{*}$, and without introducing adjustable parameters, we compute the critical temperature $T_c$ as a function of film thickness for representative noble, alkali, and alkaline-earth metals. The theory predicts that superconductivity can emerge only in selected cases and within extremely narrow thickness windows, typically centered around sub-nanometer scales ($L \sim 0.4$--$0.6$~nm), highlighting a pronounced fine-tuning requirement for confinement-induced superconductivity in good metals.  
We further discuss layered superconductor/normal-metal systems, where quantum confinement and proximity effects coexist. In these heterostructures, a substantial enhancement of the critical temperature is predicted, even when the constituent materials are non-superconducting or poor superconductors in bulk form.
\end{abstract}

\maketitle

\section{Introduction}

Tailoring the superconductivity of metals through reduced dimensionality and spatial confinement has long been explored as a route toward enhancing critical temperatures and uncovering new superconducting states \cite{Ginzburg,BuzeaRobbie,Matsuoka2026}. 
Advances in thin-film growth, interface engineering, and nanoscale fabrication \cite{tony1} have made it possible to access regimes where quantum confinement strongly modifies the electronic spectrum, leading to striking deviations from bulk behavior. 
In parallel, the application of extreme pressure has emerged as a powerful tool to induce superconductivity in elements that are non-superconducting at ambient conditions, revealing a remarkable richness of superconducting phases across the periodic table \cite{BuzeaRobbie,Matsuoka2026}.

In crystalline superconducting films, such as Pb and Al, confinement has been shown to produce pronounced, often non-monotonic, variations of the critical temperature with thickness, including enhancements above the bulk value \cite{nostro}. Confinement has also been shown to modify the penetration depth of magnetic fields \cite{forn}.
Similar size- and structure-induced superconducting phenomena have also been reported in selected elemental systems in thin-film, granular, or interfacial form, even when bulk superconductivity is absent \cite{BuzeaRobbie}. 
In the following, we refer to the \textit{ultra-thin} regime as film thicknesses in the sub-nanometer range (typically a few \AA), where confinement effects dominate the electronic structure and strongly modify the density of states near the Fermi level \cite{Zaccone_Rev}.

Early microscopic studies based on the Bogoliubov--de Gennes formalism and quantum well models demonstrated that quantum-size effects and shape resonances can induce oscillatory and enhanced superconducting critical temperatures in metallic nanofilms, nanowires, and confined geometries \cite{ShanenkoEPL2006,ShanenkoPRB2006,ShanenkoPRB2007,ShanenkoRectangular2007,ShanenkoPRL2007}, matching early experimental evidence \cite{GuoSci2004}.

These observations naturally raise a broader and more fundamental question: can quantum confinement alone induce superconductivity in elemental metals that are not superconducting in bulk, such as noble metals and several $s$-block elements?

From a conventional viewpoint, these materials are excluded from superconductivity at ambient pressure because their electron--phonon coupling is too weak to overcome Coulomb repulsion. However, recent theoretical work has demonstrated that, in the ultra-thin regime, confinement-driven reshaping of the electronic spectrum and Fermi-surface topology can significantly enhance the density of states at the Fermi level and renormalize the effective electron--phonon coupling \cite{zacbcs,noblemetal,Mg}. This raises the possibility that superconductivity may emerge as a purely confinement-induced instability, albeit under highly constrained conditions.

In this article, we provide a quantitative perspective on this problem by formulating a confinement-generalized isotropic Eliashberg theory for metallic thin films. In this framework, quantum confinement induces an energy-dependent normal density of states and generates a thickness dependence of key material parameters, including the Fermi energy $E_F$, the electron--phonon coupling constant $\lambda$, and the Coulomb pseudopotential $\mu^{*}$. Importantly, the theory contains no adjustable parameters and smoothly recovers the standard bulk Eliashberg formalism in the limit of large thickness.

By solving the resulting equations numerically using ab-initio or experimentally determined electron--phonon spectral functions, we assess the emergence of superconductivity in noble metals, alkali metals, and alkaline-earth metals. We show that confinement-induced superconductivity is possible only in selected cases and only within extremely narrow thickness windows, highlighting a strong fine-tuning requirement. Finally, we extend the discussion to superconductor/normal-metal multilayers, where quantum confinement and proximity effects coexist, and demonstrate that such engineered heterostructures may provide a more robust route to enhanced superconducting critical temperatures.

\section{Theoretical framework}

\subsection{Electronic structure of metallic thin films}

Metallic thin films provide a paradigmatic platform where spatial confinement profoundly alters the electronic structure relative to the bulk. Within a free-electron description, deviations from ideal hard-wall boundary conditions inevitably arise due to surface roughness and imperfect confinement. Consequently, the electronic wavefunctions do not vanish sharply at the film boundaries, and the out-of-plane momentum component cannot be treated as a strictly quantized good quantum number. 
For example, nanometric-thin films assembled by cluster beam deposition exhibit complex conduction behavior due to high densities of grain boundaries and defects, emphasizing that surface disorder leads to strong deviations from ideal quantized electronic sub-bands \cite{MiriglianoMilani2021}. A similar situation is encountered in ultra-thin semiconductor films \cite{ZacconePRMaterials9}.

Indeed, a pervasive practical complication in ultra–thin metallic films is the presence of surface roughness and structural disorder at the film boundaries, which is unavoidable in realistic experimental systems. In ideal hard–wall quantum well models, the transverse momentum \(k_z\) is treated as a precisely quantized good quantum number, leading to discrete sub-bands associated with standing waves across the film thickness. However, in the presence of surface disorder and imperfect confinement, these standing waves are perturbed and broadened, and strict quantization of \(k_z\) no longer holds \cite{KassubekHaug,FeenstraSurface}, an effect observed also with confined phonons \cite{NatureComm}. Disorder at the interfaces mixes transverse momentum states and smears discrete levels into a quasi–continuum, suppressing coherent oscillatory behavior expected in perfectly ordered films. The confinement model adopted here does not assume exact \(k_z\) quantization but instead describes the electronic spectrum through an effective, thickness–dependent density of states that evolves continuously with film thickness. In this way, the model naturally incorporates the effects of ubiquitous surface disorder on the electronic structure and provides a more realistic description of the phase space available for electron pairing in ultra–thin films.

Nevertheless, confinement strongly reshapes the topology of the available phase space for itinerant electrons, as discussed in the following section.

\subsubsection{Confinement-induced Fermi surface modification}

The primary effect of reduced thickness is a suppression of electronic states with momenta close to the confinement direction. In momentum space, this manifests as the removal of two symmetric regions along the out-of-plane axis, which can be viewed as hole-like pockets carved out of the bulk Fermi sphere. As a result, the accessible volume in $k$ space is reduced compared to the three-dimensional case.

\begin{figure}
    \centering
    \includegraphics[width=\linewidth]{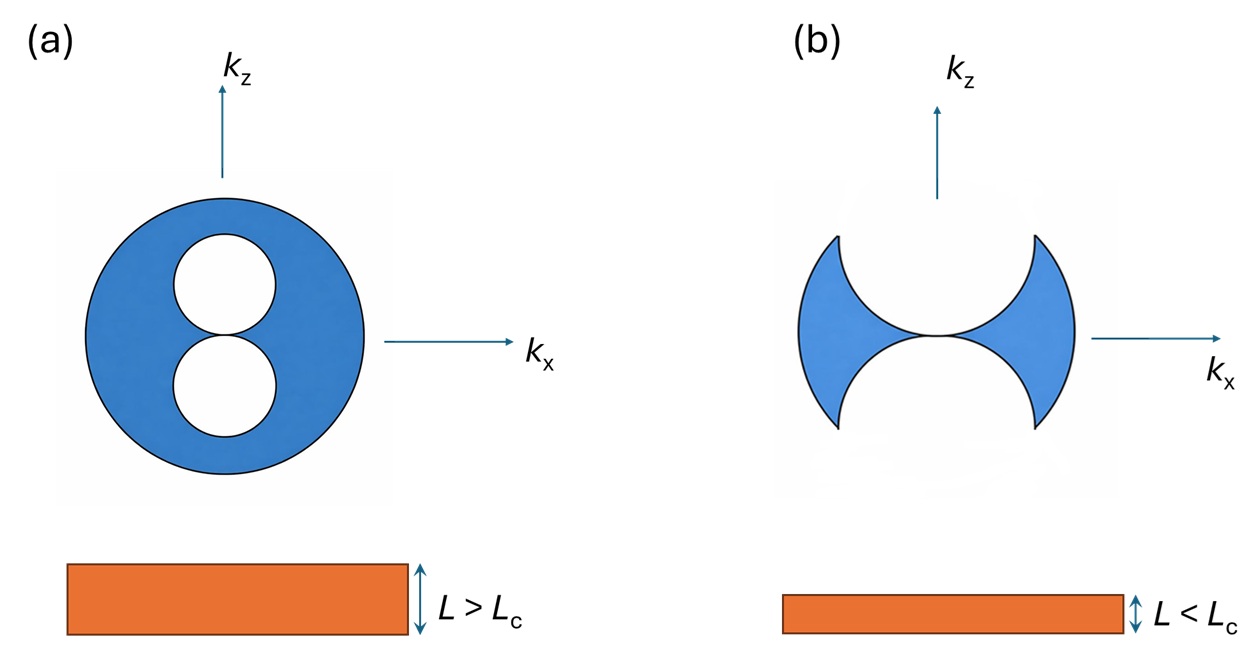}
    \caption{Schematic illustration of the confinement-induced reconstruction of the Fermi surface in a metallic thin film. 
    (Left) For weak confinement, the bulk Fermi sphere is partially depleted by two symmetric hole-like regions along the confinement direction, corresponding to suppressed electronic states.
    (Right) Below a critical thickness, the overlap of these excluded regions leads to a topological transformation of the Fermi surface into a non-simply-connected geometry. 
    The orange bars indicate the film thickness $L$, highlighting the role of spatial confinement in driving the topological crossover.}
    \label{fig:FS_topology}
\end{figure}

When the film thickness is decreased below a critical value determined by the carrier density, this geometric modification evolves into a qualitative change in the topology of the Fermi surface. Specifically, the initially simply connected spherical surface undergoes a confinement-driven topological crossover into a nontrivial configuration characterized by a different homotopy class. A schematic representation of this confinement-induced reconstruction is shown in Fig.~\ref{fig:FS_topology}. This transition, its geometric interpretation, and its implications for the electronic phase space are discussed in detail in Ref.~\cite{zacbcs}.

\subsubsection{Density of states and crossover energy scale}

The reconstruction of the Fermi surface directly impacts the electronic density of states (DOS). A characteristic confinement energy scale emerges,
\begin{equation}
\epsilon^{*} = \frac{2\pi^{2}\hbar^{2}}{mL^{2}},
\end{equation}
which separates two distinct regimes. At energies below $\epsilon^{*}$, the DOS exhibits a linear dependence on energy, reflecting the effective reduction of dimensionality induced by confinement. At higher energies, the standard square-root dependence of a three-dimensional free-electron gas is recovered. This crossover behavior and its microscopic origin are derived analytically in Ref.~\cite{zacbcs}.

\subsubsection{Thickness dependence of the Fermi energy}

At zero temperature, the Fermi energy is determined by particle-number conservation. Owing to the thickness dependence of the DOS, the Fermi energy acquires an explicit dependence on the film thickness. The two confinement regimes are separated by a critical thickness
\begin{equation}
L_{c} = \left( \frac{2\pi}{n} \right)^{1/3},
\end{equation}
where $n$ denotes the free-carrier density.

In the strong-confinement regime ($L < L_{c}$), where the linear DOS extends up to the Fermi level, the Fermi energy follows the scaling
\begin{equation}
\epsilon_{F} \sim \frac{\hbar^{2}}{m}\left( \frac{n}{L} \right)^{1/2}.
\end{equation}
For thicker films, the Fermi energy approaches its bulk value with confinement-induced corrections. The full piecewise expression and its connection to earlier hard-wall models are presented in Ref.~\cite{zacbcs}.

\subsection{Eliashberg theory with confinement}

The microscopic description of phonon-mediated superconductivity beyond the weak-coupling BCS approximation is provided by Eliashberg theory, in which the pairing interaction is fully characterized by the electron--phonon spectral function $\alpha^{2}F(\Omega)$ and the residual Coulomb repulsion is accounted for by the Coulomb pseudopotential $\mu^{*}$ \cite{ummarev,revcarbi}. In the bulk, this framework has proven quantitatively accurate for a wide range of conventional superconductors.

Here we generalize the isotropic, one-band Eliashberg formalism to ultra-thin metallic films, where quantum confinement alters the electronic structure. In the limit of large thickness, the confined theory smoothly reduces to the standard bulk Eliashberg equations. The key physical effect of confinement is that the normal density of states (NDOS) near the Fermi level can no longer be approximated as a constant, and the electronic bandwidth becomes thickness dependent.

\subsubsection{Standard Eliashberg equations}

We briefly recall the standard formulation on the imaginary axis. In the isotropic one-band approximation and assuming Migdal’s theorem holds \cite{ummaMig}, the superconducting state is described by the gap function $\Delta(i\omega_n)$ and the mass-renormalization function $Z(i\omega_n)$, which satisfy
\begin{equation}
\Delta(i\omega_n)Z(i\omega_n)=\pi T\sum_{\omega_{n'}} 
\frac{\Delta(i\omega_{n'})}{\sqrt{\omega_{n'}^2+\Delta^{2}(i\omega_{n'})}}
\Big[ \lambda(i\omega_{n'}-i\omega_n)-\mu^{*}(\omega_{c})\theta(\omega_{c}-|\omega_{n'}|)\Big],
\end{equation}
\begin{equation}
Z(i\omega_n)=1+\frac{\pi T}{\omega_n}\sum_{\omega_{n'}}
\frac{\omega_{n'}}{\sqrt{\omega_{n'}^2+\Delta^{2}(i\omega_{n'})}}
\lambda(i\omega_{n'}-i\omega_n).
\end{equation}

Here $\omega_n$ are fermionic Matsubara frequencies and $\omega_c$ is a high-energy cutoff exceeding the maximum phonon frequency. The pairing kernel is determined by the electron--phonon spectral function,
\begin{equation}
\lambda(i\omega_{n'}-i\omega_n)=2\int_0^\infty d\Omega\,
\frac{\Omega\,\alpha^2F(\Omega)}{\Omega^2+(\omega_{n'}-\omega_n)^2},
\end{equation}
while the dimensionless coupling constant is $\lambda=2\int_0^\infty d\Omega\,\alpha^2F(\Omega)/\Omega$.
These coupled nonlinear equations are solved numerically, and the critical temperature $T_c$ is determined from the temperature at which a nontrivial solution for $\Delta$ emerges.

\subsubsection{Confinement-induced modifications}

Quantum confinement requires relaxing two approximations implicit in the bulk theory: the infinite-band approximation and the assumption of a constant NDOS near the Fermi level \cite{ummachi,Allen}. In the most general case this introduces additional self-energy terms; however, when the NDOS is symmetric with respect to the Fermi energy, $N(\varepsilon)=N(-\varepsilon)$, the theory simplifies and retains the same two self-consistency equations for $Z$ and $\Delta$, albeit with modified kernels \cite{carbin1,carbin2}.

In confined metallic films, the NDOS acquires a thickness-dependent, energy-dependent form governed by a confinement energy scale $\varepsilon^{*}$
which separates a low-energy linear regime from the high-energy three-dimensional square-root behavior \cite{zacbcs}. Two distinct confinement regimes arise, depending on whether the film thickness $L$ is larger or smaller than the critical thickness $L_c$ defined by the carrier density.
The Eliashberg equations  generalized with confinement effect are \cite{nostro}:
\begin{equation}
\begin{split}
Z(i\omega_n)= 1+\frac{\pi T}{\omega_n}\sum_{\omega_{n'}} \frac{\omega_{n'}}{\sqrt{\omega_{n'}^2+\Delta^{2}(i\omega_{n'})}}[\frac{N(i\omega_{n'})+N(-i\omega_{n'})}{2}]\times
\\
\times\lambda (i\omega_{n'}-i\omega_n)\frac{2}{\pi}\arctan(\frac{W}{2Z(i\omega_{n'})\sqrt{\omega_{n'}^{2}+\Delta^{2}(i\omega_{n'})}})
\end{split}
\end{equation}

\begin{equation}
\begin{split}
\Delta(i\omega_n)Z(i\omega_n)=\pi T\sum_{\omega_{n'}} \frac{\Delta(i\omega_{n'})}{\sqrt{\omega_{n'}^2+\Delta^{2}(i\omega_{n'})}}[\frac{N(i\omega_{n'})+N(-i\omega_{n'})}{2}]\times
\\
\times\big[\lambda(i\omega_{n'}-i\omega_n)-\mu^{*}(\omega_c)\theta(\omega_c-|\omega_{n'}|)\big]\frac{2}{\pi} \arctan(\frac{W}{2Z(i\omega_{n'})\sqrt{\omega_{n'}^{2}+\Delta^{2}(i\omega_{n'})}})
\end{split}
\end{equation}
where $N(\pm i\omega_{n'})=N(\pm Z(i\omega_{n'})\sqrt{(\omega_{n'})^{2}+\Delta^{2}(i\omega_{n'})})$ and the bandwidth $W$ is equal to half the Fermi energy, $E_{F}/2$.

We summarize the key consequences of confinement for the Eliashberg framework:
\begin{itemize}
\item[(i)] the NDOS becomes an explicit function of energy, $N(\varepsilon)$, reflecting the confinement-induced reconstruction of the electronic spectrum;
\item[(ii)] the Fermi energy is renormalized according to $E_F=C(L)^2E_{F,\mathrm{bulk}}$, leading to a thickness-dependent electronic bandwidth;
\item[(iii)] the effective electron--phonon coupling is enhanced as $\lambda=C(L)\lambda_{\mathrm{bulk}}$;
\item[(iv)] the Coulomb pseudopotential acquires a corresponding thickness dependence,
\begin{equation}
\mu^{*}(L)=\frac{C(L)\mu_{\mathrm{bulk}}}{1+\mu_{\mathrm{bulk}}\ln[E_F(L)/\omega_c]}.
\end{equation}
\end{itemize}

Here the dimensionless factor $C(L)$ is fully determined by the confinement geometry and carrier density, and no adjustable parameters are introduced. Explicit expressions for $N(\varepsilon)$ in the different confinement regimes, as well as the resulting piecewise forms of $E_F(L)$ and $C(L)$, are given in Ref.~\cite{zacbcs}.

\subsubsection{Numerical implementation}

In the numerical solution of the Eliashberg equations, energies are measured relative to the Fermi level, which is taken as the zero of energy. The NDOS is correspondingly rescaled to ensure continuity at $\varepsilon=\varepsilon^{*}$. This formulation allows for stable numerical convergence and enables a direct, parameter-free computation of $T_c$ as a function of film thickness across the entire confinement regime.

\begin{figure*}[ht]
	\centerline{\includegraphics[width=1\textwidth]{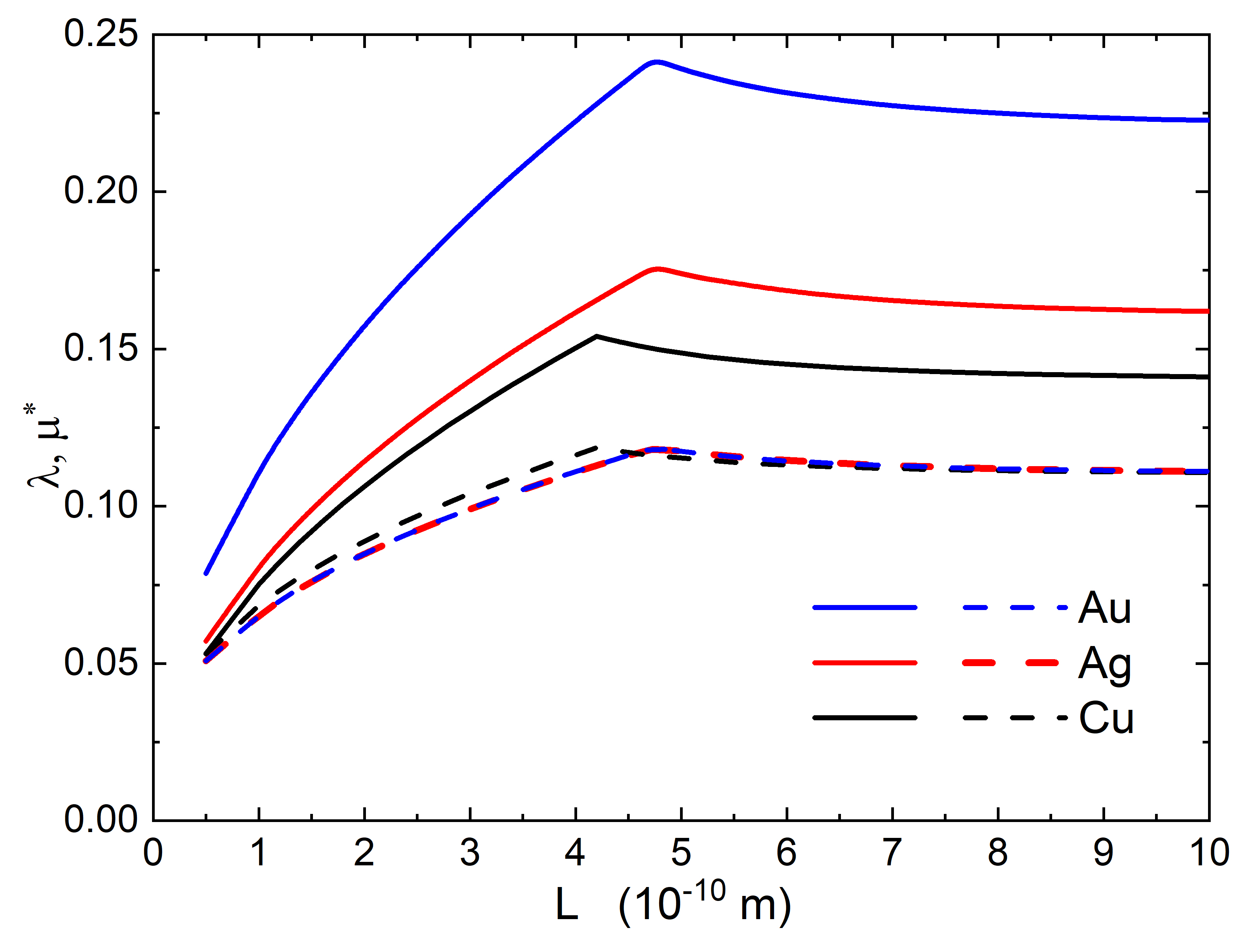}}
\caption{Noble metals. Physical parameters used in the theory for films of different materials:
Cu ($\lambda$ (black solid line) and $\mu^{*}$ (black dashes line)),
Ag ($\lambda$ (red solid line) and $\mu^{*}$ (red dashes line)) and
Au ($\lambda$ (dark blue solid line) and $\mu^{*}$ (dark blue dashes line)).
All parameters are plotted as a function of the film thickness $L$.
}\label{fig:noble_params}
\end{figure*}

\begin{figure*}[t!]	\centerline{\includegraphics[width=1\textwidth]{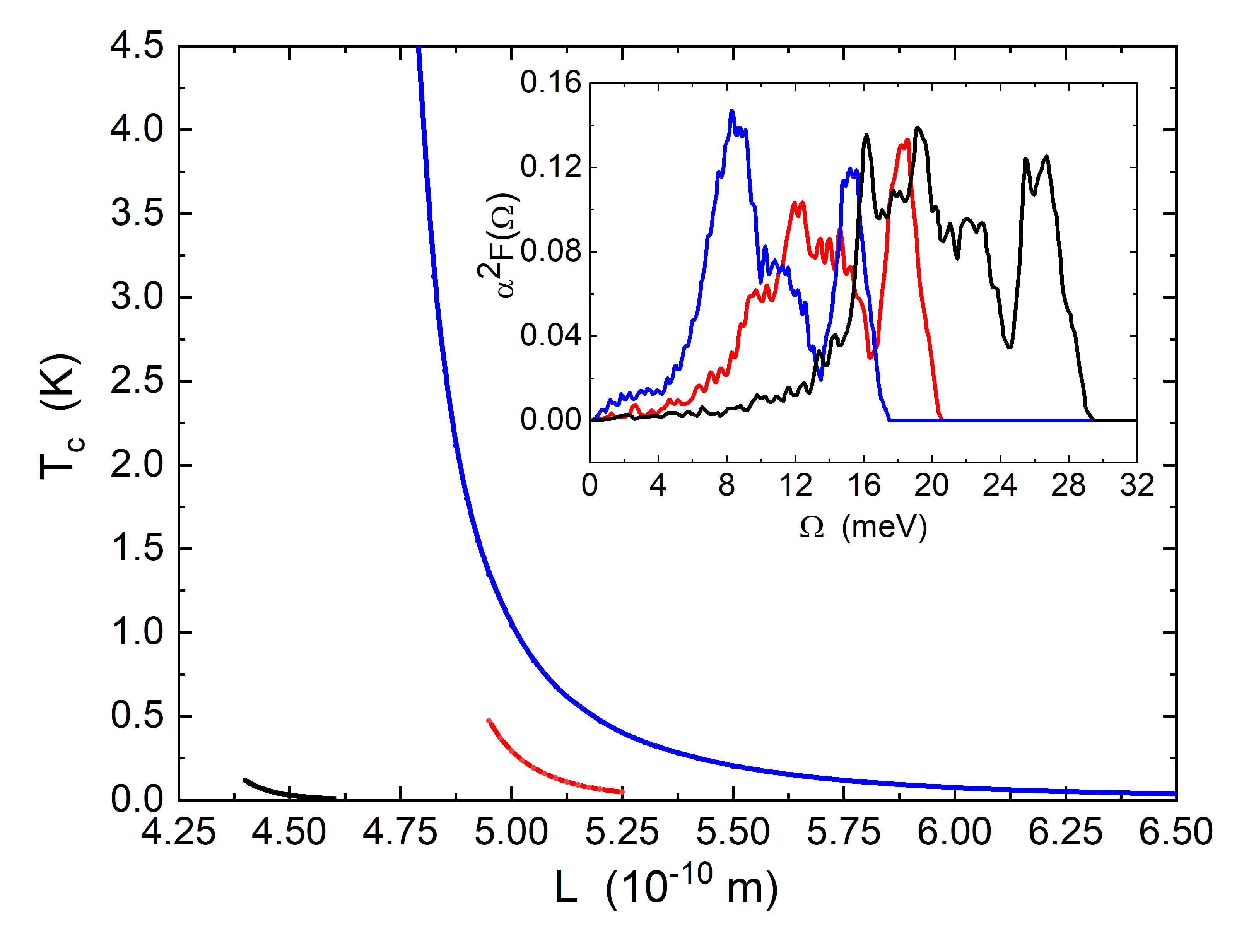}}
\caption{Noble metals. Critical temperature $T_c$ versus film thickness $L$: full solid line represent the numerical solutions of Eliashberg equations.
Cu (black solid line), Ag (red solid line) and Au (dark blue solid line).
In the inset, the Eliashberg electron-phonon spectral function of these elements are shown: Cu (black solid line), Ag (red solid line) and Au (dark blue solid line).}\label{fig:noble_tc}
\end{figure*}

\begin{figure*}[ht]
	\centerline{\includegraphics[width=1\textwidth]{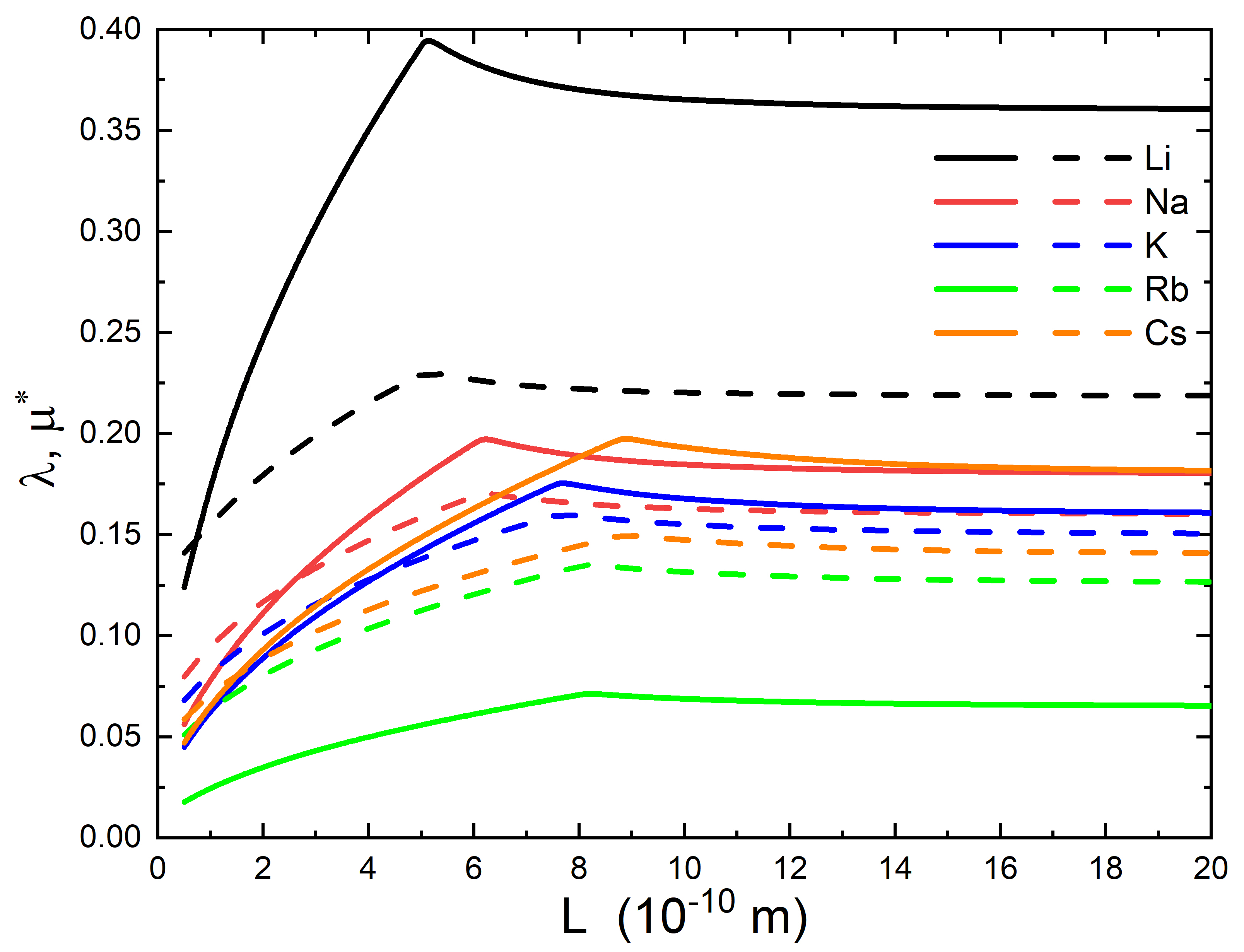}}
\caption{Alkali metals. Physical parameters used in the theory for films of different materials:
Li ($\lambda$ (black solid line) and $\mu^{*}$ (black dashes line)),
Na ($\lambda$ (red solid line) and $\mu^{*}$ (red dashes line)) and
K ($\lambda$ (blue solid line) and $\mu^{*}$ (blue dashes line)).
Rb ($\lambda$ (green solid line) and $\mu^{*}$ (green dashes line)) and
Cs ($\lambda$ (orange solid line) and $\mu^{*}$ (orange dashes line)).
All parameters are plotted as a function of the film thickness $L$.
}\label{fig:alkali_params}
\end{figure*}

\begin{figure*}[t!]	\centerline{\includegraphics[width=1\textwidth]{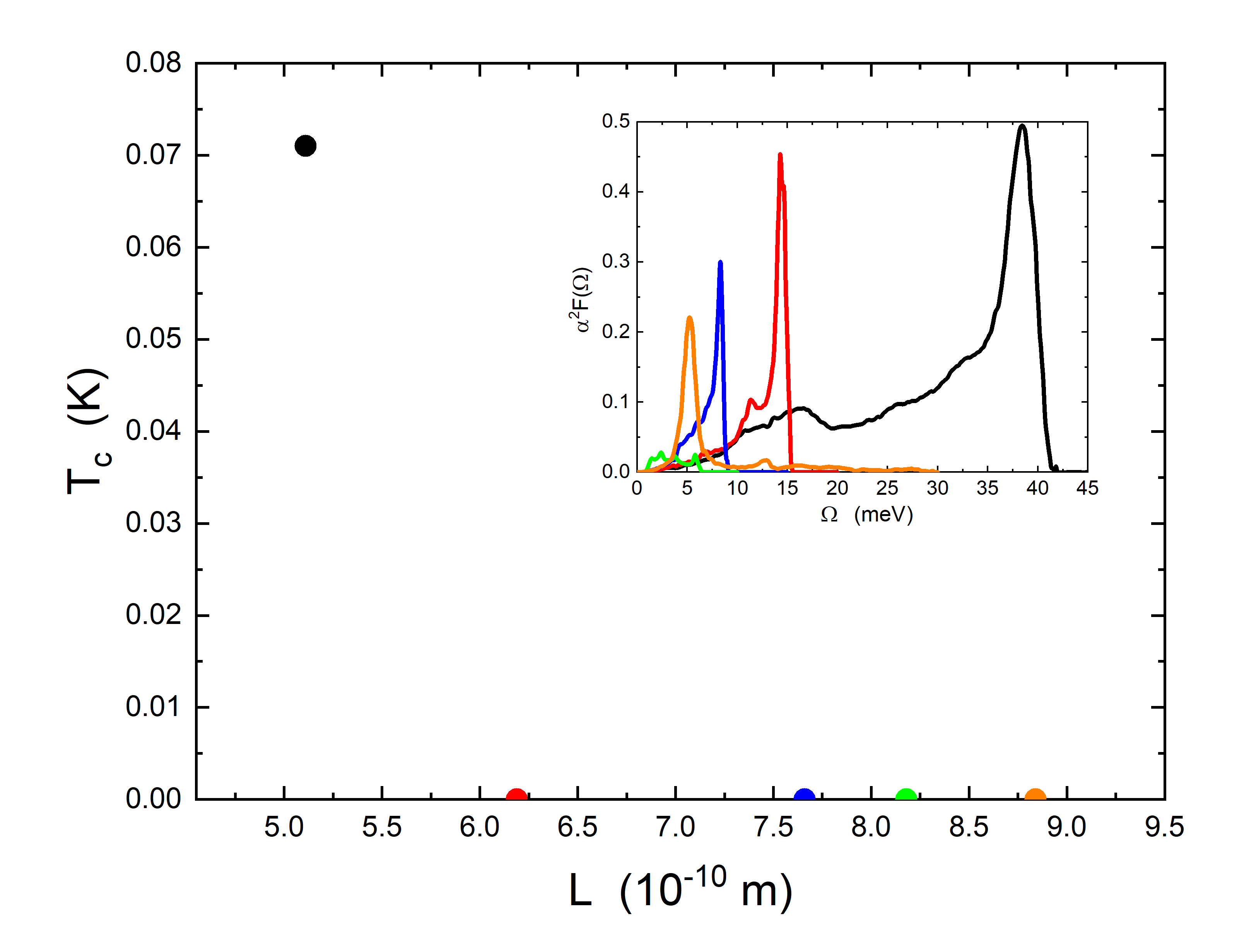}}
\caption{Alkali metals. Critical temperature $T_c$ versus film thickness $L$: full circles represent the numerical solutions of Eliashberg equations for obtaining the maximum $T_c$.
Li (black full circle), Na (red full circle), K (green full circle), Rb (blue full circle) and Cs (orange full circle).
In the inset, the Eliashberg electron-phonon spectral function of these elements are shown: Li (black solid line), Na (red solid line), K (green solid line), Rb (blue solid line) and Cs (orange solid line).}\label{fig:alkali_tc}
\end{figure*}

\begin{figure*}[ht]
	\centerline{\includegraphics[width=1\textwidth]{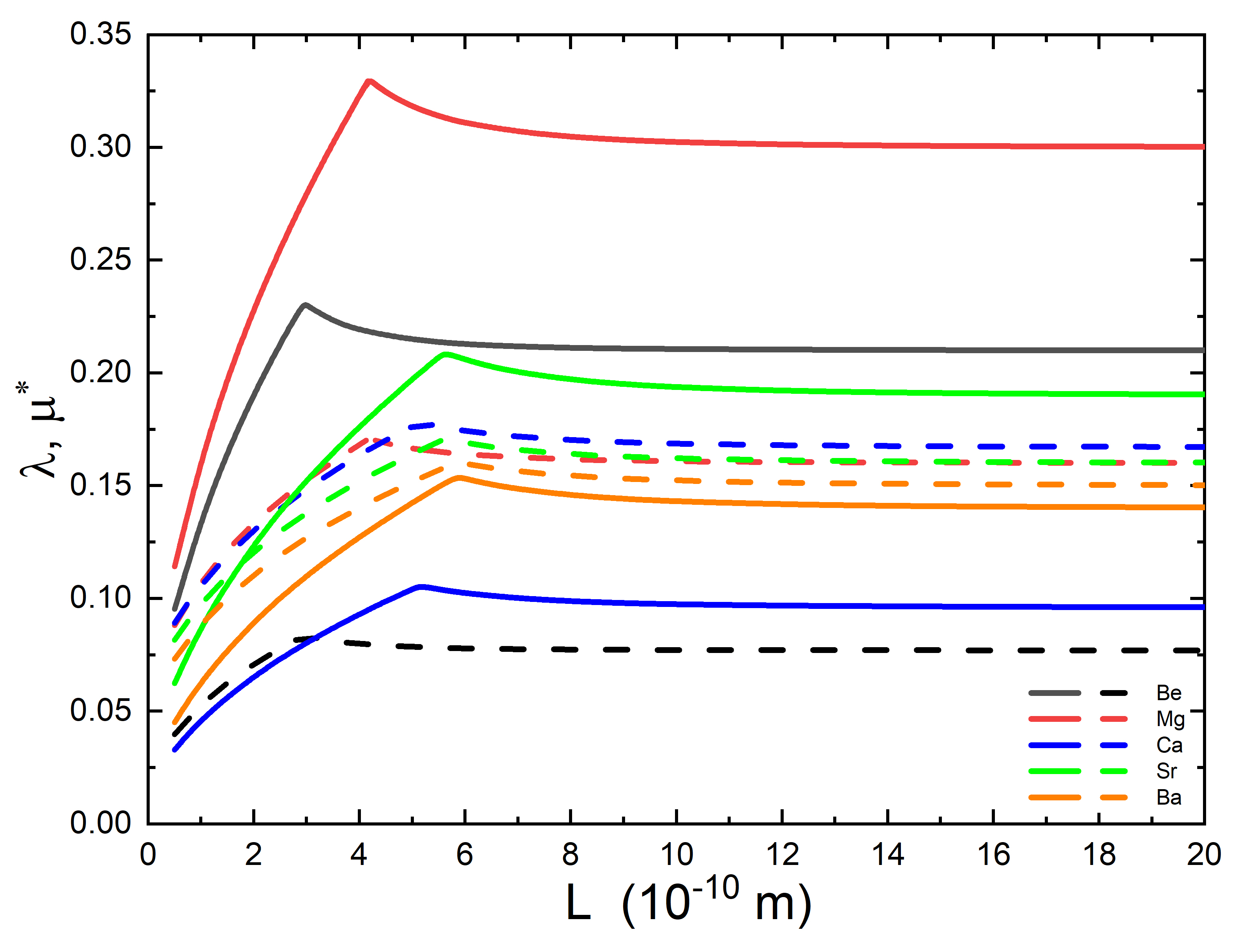}}
\caption{Alkaline-earth metals. Physical parameters used in the theory for films of different materials:
Be ($\lambda$ (black solid line) and $\mu^{*}$ (black dashes line)),
Mg ($\lambda$ (red solid line) and $\mu^{*}$ (red dashes line)) and
Ca ($\lambda$ (blue solid line) and $\mu^{*}$ (blue dashes line)).
Sr ($\lambda$ (green solid line) and $\mu^{*}$ (green dashes line)) and
Ba ($\lambda$ (orange solid line) and $\mu^{*}$ (orange dashes line)).
All parameters are plotted as a function of the film thickness $L$.
}\label{fig:alkaline_params}
\end{figure*}

\begin{figure*}[t!]	\centerline{\includegraphics[width=1\textwidth]{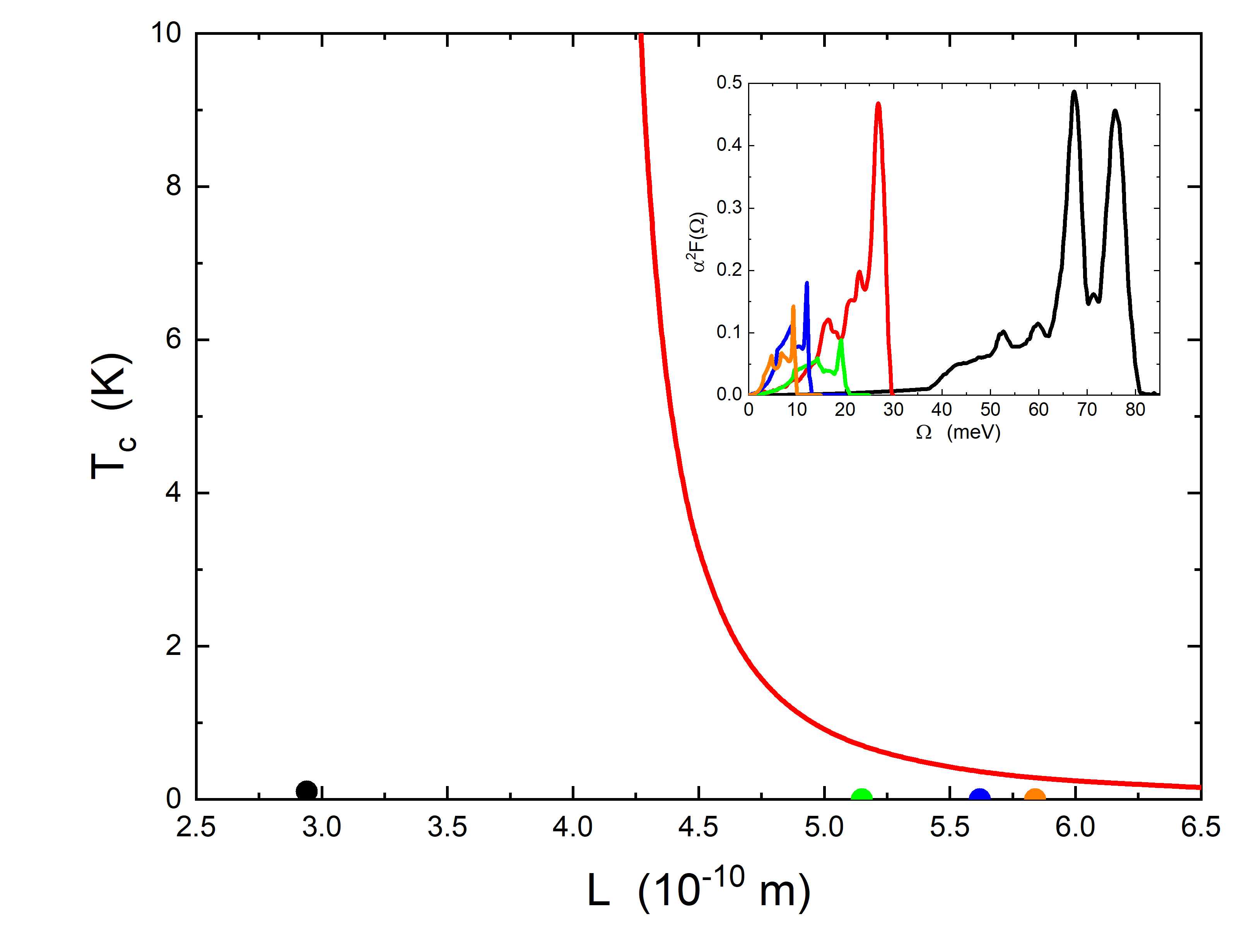}}
\caption{Alkaline-earth metals. Critical temperature $T_c$ versus film thickness $L$: full solid line represent the numerical solutions of Eliashberg equations.
Be (black full circle), Mg (red solid line), Ca (green full circle), Sr (blue full circle) and Ba (orange full circle).
In the inset, the Eliashberg electron-phonon spectral function of these elements are shown: Be (black solid line), Mg (red solid line), Ca (green solid line), Sr (blue solid line) and Ba (orange solid line).}\label{fig:alkaline_tc}
\end{figure*}


\begin{table}[t]
\centering
\small
\setlength{\tabcolsep}{3.5pt}
\renewcommand{\arraystretch}{1.15}
\resizebox{\textwidth}{!}{%
\begin{tabular}{|c|c|c|c|c|c|c|c|c|c|}
  \hline
  metal & $\lambda_{bulk}$ & $\mu^{*}(\omega_c)$ & $\omega_c$ (meV) & $n_0$ ($10^{28}$ m$^{-3}$) & $E_{F,bulk}$ (eV) & $T_{c,bulk}$ (K) & $L_c$ (\AA) & $T_{c,m}$ (K) & $\omega_m$ (meV) \\
  \hline
  Cu & 0.140 & 0.1100 & 90 & 8.47 & 7.00 & 0 & 4.20 & 0.118 & 100 \\
  Ag & 0.160 & 0.1100 & 75 & 5.86 & 5.49 & 0 & 4.75 & 0.471 & 80 \\
  Au & 0.220 & 0.1100 & 55 & 5.90 & 5.54 & 0 & 4.74 & 4.515 & 60 \\
  Li & 0.360 & 0.2186 & 126 & 4.70 & 4.74 & 0.0004 & 5.11 & 0.071 & 130 \\
  Na & 0.180 & 0.1500 & 46 & 2.65 & 3.24 & 0 & 6.19 & $< 0.005$ & 30 \\
  K  & 0.160 & 0.1500 & 28 & 1.40 & 2.12 & 0 & 7.66 & $< 0.005$ & 48 \\
  Rb & 0.065 & 0.1260 & 21 & 1.15 & 1.99 & 0 & 8.18 & 0 & 22 \\
  Cs & 0.180 & 0.1400 & 90 & 0.91 & 1.59 & 0 & 8.84 & $< 0.025$ & 92 \\
  Be & 0.210 & 0.0770 & 255 & 24.70 & 14.30 & 0.026 & 2.94 & 0.1 & 260 \\
  Mg & 0.300 & 0.1600 & 90 & 8.61 & 7.08 & 0 & 4.18 & 19.557 & 100 \\
  Ca & 0.096 & 0.1670 & 66 & 4.61 & 4.69 & 0 & 5.15 & 0 & 70 \\
  Sr & 0.190 & 0.1600 & 39 & 3.55 & 3.93 & 0 & 5.62 & $< 0.005$ & 40 \\
  Ba & 0.140 & 0.1500 & 30 & 3.15 & 3.64 & 0 & 5.84 & 0 & 32 \\
  \hline
\end{tabular}%
}
\caption{Input bulk parameters and predicted maximum critical temperature $T_{c,m}$ under confinement.}
\label{tab:params}
\end{table}

\section{Prediction of critical temperature}

In the preceding section we discussed how the Eliashberg formalism is modified when the normal density of states (NDOS) acquires an energy dependence due to quantum confinement. When the NDOS is symmetric with respect to the Fermi level, the theoretical description simplifies and the superconducting state can be obtained by solving only two coupled Eliashberg equations. In contrast, if the NDOS is asymmetric, the formalism requires an additional equation together with a self-consistency condition, leading to a more involved numerical treatment \cite{ummachi}. 

In practice, the influence of the NDOS asymmetry on the superconducting properties is usually a higher-order correction, and for most materials it does not qualitatively alter the resulting critical temperature \cite{carbin1}. The formulation adopted here therefore provides a reliable description while remaining computationally tractable. Importantly, the theory contains no adjustable parameters: quantum confinement modifies the electronic structure in a well-defined manner and leads to a renormalization of the electron--phonon coupling constant according to $\lambda = C\,\lambda_{\mathrm{bulk}}$, where the confinement factor satisfies $1 \leq C \leq 1.1$. Since the critical temperature generally increases with increasing $\lambda$ \cite{ummarev}, the maximum value of $T_c$ is expected to occur near the characteristic thickness $L=L_c$, where the enhancement factor $C$ reaches its maximum.

In some cases the numerical solution exactly at $L=L_c$ becomes difficult due to convergence issues associated with the extremely small energy scales involved. When this occurs, the critical temperature is evaluated for thickness values slightly larger than $L_c$, which nevertheless capture the maximum of the superconducting instability with good accuracy. In the following sections we apply this framework to several classes of elemental metals, namely noble metals (Cu, Ag, Au), alkali metals (Li, Na, K, Rb, Cs), and alkaline-earth metals (Be, Mg, Ca, Sr, Ba), in order to assess whether quantum confinement can induce superconductivity in systems that are non-superconducting in bulk form.

Within this framework, quantum confinement leads to a systematic renormalization of the effective electron--phonon interaction. In particular, the coupling constant scales as $\lambda = C\,\lambda_{\mathrm{bulk}}$, where the dimensionless factor $C$ depends on the film thickness and typically varies within the interval $1 \leq C \leq 1.1$. Because the superconducting critical temperature generally increases with stronger electron--phonon coupling \cite{ummarev}, the enhancement of $\lambda$ induced by confinement naturally produces a maximum of $T_c$ near the characteristic thickness $L=L_c$, where the factor $C$ attains its largest value.

In some situations the direct numerical evaluation of $T_c$ exactly at $L=L_c$ becomes difficult due to convergence issues associated with the extremely small energy scales involved in the Eliashberg equations. In these cases we evaluate the critical temperature for thicknesses slightly larger than $L_c$, which nevertheless provide a reliable estimate of the maximum superconducting instability.

It should be noted that lithium and beryllium already exhibit superconductivity in bulk form, although only at extremely low temperatures ($T_c=4\times10^{-4}$~K and $T_c=0.026$~K, respectively). Because these values are experimentally difficult to access, it is of practical interest to explore whether quantum confinement can significantly enhance their critical temperatures.

\section{Copper, silver, and gold}

It is well established that the noble metals Cu, Ag, and Au are not superconducting in bulk form at ambient pressure, owing to their weak electron--phonon coupling ($\lambda_{\mathrm{bulk}}<0.25$), which is insufficient to overcome Coulomb repulsion. However, when these materials are confined to ultra-thin films with thicknesses close to the critical confinement length $L_c$—typically of the order of $5\,\text{\AA}$ ($\sim0.5$~nm)—our calculations show that the effective electron--phonon interaction can be enhanced. As a result, superconductivity may emerge within a very narrow thickness window, as discussed in Ref.~\cite{noblemetal}.

\subsection*{Copper}

We begin with copper and examine the behavior of the relevant physical parameters in the vicinity of the critical thickness $L_c$.  
Figure~\ref{fig:noble_params} shows the thickness dependence of the electron--phonon coupling constant $\lambda$ and the Coulomb pseudopotential $\mu^{*}$ for Cu (black curves). The bulk electron--phonon spectral function $\alpha^2F(\Omega)$, corresponding to $\lambda_{\mathrm{bulk}}=0.14$, is shown in the inset of Fig.~\ref{fig:noble_tc} (black curve) \cite{GIRI}. The bulk Coulomb pseudopotential is $\mu^{*}_{\mathrm{bulk}}(\omega_c)=0.11$, with a cutoff energy $\omega_c=90$~meV and a maximum phonon energy $\omega_{\max}=100$~meV. The bulk Fermi energy and carrier density are $E_{F,\mathrm{bulk}}=7000$~meV and $n_0=8.47\times10^{28}$~m$^{-3}$, respectively \cite{Mermin}.

Solving the confinement-modified Eliashberg equations, we find that for a film thickness $L=4.40\,\text{\AA}$—close to the critical value $L_c=4.20\,\text{\AA}$—copper becomes superconducting with a maximum critical temperature $T_c=0.118$~K, as shown in Fig.~\ref{fig:noble_tc}. The thickness interval over which superconductivity is stabilized is extremely narrow, of the order of $\sim0.25\,\text{\AA}$, indicating a pronounced fine-tuning requirement. For thicknesses even closer to $L_c$, numerical convergence becomes increasingly difficult and reliable solutions cannot be obtained.

\subsection*{Silver}

The corresponding results for silver are shown in Figs.~\ref{fig:noble_params} and~\ref{fig:noble_tc} (red curves). The bulk electron--phonon spectral function of Ag, with $\lambda_{\mathrm{bulk}}=0.16$, is displayed in the inset of Fig.~\ref{fig:noble_tc} (red curve) \cite{GIRI}. The bulk Coulomb pseudopotential is $\mu^{*}_{\mathrm{bulk}}(\omega_c)=0.11$, with $\omega_c=75$~meV and $\omega_{\max}=80$~meV. The bulk Fermi energy and carrier density are $E_{F,\mathrm{bulk}}=5490$~meV and $n_0=5.86\times10^{28}$~m$^{-3}$, yielding a critical thickness $L_c=4.75\,\text{\AA}$ \cite{Mermin}.

As seen in Fig.~\ref{fig:noble_params}, both $\lambda$ (solid red line) and $\mu^{*}$ (dashed red line) increase slightly in the vicinity of $L_c$. To determine whether this enhancement suffices to induce superconductivity, we solve the Eliashberg equations and compute the critical temperature. The resulting $T_c(L)$ curve is shown in Fig.~\ref{fig:noble_tc} (red curve). We find that at $L=4.95\,\text{\AA}$, close to $L_c$, silver becomes superconducting with $T_c=0.472$~K. As in the copper case, the superconducting phase exists only within a very narrow thickness window of approximately $0.30\,\text{\AA}$. This behavior contrasts sharply with conventional superconductors such as Pb and Al, where confinement-induced effects persist over much broader thickness ranges \cite{nostro}.

\subsection*{Gold}

Finally, we consider gold. The thickness dependence of $\lambda$ and $\mu^{*}$ for Au is shown in Fig.~\ref{fig:noble_params} (green curves), while the corresponding $T_c(L)$ is reported in Fig.~\ref{fig:noble_tc} (green curve). The bulk electron--phonon spectral function of Au, characterized by $\lambda_{\mathrm{bulk}}=0.22$, is shown in the inset of Fig.~\ref{fig:noble_tc} (green curve) \cite{GIRI}. The bulk Coulomb pseudopotential is $\mu^{*}_{\mathrm{bulk}}(\omega_c)=0.11$, with $\omega_c=55$~meV and $\omega_{\max}=60$~meV. The bulk Fermi energy and carrier density are $E_{F,\mathrm{bulk}}=5530$~meV and $n_0=5.90\times10^{28}$~m$^{-3}$, leading to a critical thickness $L_c=4.74\,\text{\AA}$ \cite{Mermin}.

For a film thickness $L=4.79\,\text{\AA}$, very close to $L_c$, gold becomes superconducting with a remarkably high critical temperature $T_c=4.515$~K, as shown in Fig.~\ref{fig:noble_tc}. The thickness range over which superconductivity is observed is somewhat broader than in Cu and Ag, extending over approximately $1.75\,\text{\AA}$, yet it remains strongly localized around the critical confinement length. Moving away from $L_c$, the critical temperature drops rapidly to very small values, which are computationally inaccessible due to the increasing difficulty in achieving numerical convergence.

%
\section{Lithium, sodium, potassium, rubidium, and cesium}

With the exception of lithium, the alkali metals are not superconducting at ambient pressure. Lithium itself exhibits superconductivity only at an extremely low critical temperature ($T_c \simeq 4\times10^{-4}$~K). It is therefore natural to investigate whether quantum confinement in ultra-thin films can enhance superconductivity in this class of materials.

\subsection*{Lithium}

We begin with lithium. The thickness dependence of the electron--phonon coupling constant $\lambda$ and the Coulomb pseudopotential $\mu^{*}$ used in the calculations is shown in Fig.~\ref{fig:alkali_params} (black curves). The bulk electron--phonon spectral function $\alpha^2F(\Omega)$, corresponding to $\lambda_{\mathrm{bulk}}=0.36$, is displayed in the inset of Fig.~\ref{fig:alkali_tc} (black curve) \cite{a2fLithium}. The Coulomb pseudopotential is fixed to reproduce the experimental bulk critical temperature, yielding $\mu^{*}_{\mathrm{bulk}}(\omega_c)=0.2186$, with a cutoff energy $\omega_c=126$~meV and a maximum phonon energy $\omega_{\max}=130$~meV. The bulk Fermi energy and carrier density are $E_{F,\mathrm{bulk}}=4740$~meV and $n_0=4.7\times10^{28}$~m$^{-3}$, respectively \cite{Mermin}, leading to a critical thickness $L_c=5.11\,\text{\AA}$.

Solving the confinement-modified Eliashberg equations, we find that at $L=L_c$ the critical temperature is enhanced to a maximum value of $T_c=0.072$~K, as shown in Fig.~\ref{fig:alkali_tc}. Although this represents a substantial increase compared to the bulk value, the resulting $T_c$ remains very low and confined to a narrow thickness range around $L_c$.
Of course lithium is a bulk superconductor, so it should have a continuous crossover of $T_c$ from the bulk to the ultrathin-film limit. The corresponding curve is missing because we didn't have enough computing capabilities for calculating such low critical temperatures, which required the addition of a huge number of Matsubara frequencies. 

\subsection*{Sodium}

The corresponding results for sodium are shown in Fig.~\ref{fig:alkali_params} (red curves) and Fig.~\ref{fig:alkali_tc}. The bulk electron--phonon spectral function of Na, with $\lambda_{\mathrm{bulk}}=0.18$, is reported in the inset of Fig.~\ref{fig:alkali_tc} (red curve) \cite{a2fSodium}. The bulk Coulomb pseudopotential is $\mu^{*}_{\mathrm{bulk}}(\omega_c)=0.16$ \cite{muSodium}, with $\omega_c=46$~meV and $\omega_{\max}=48$~meV. The bulk Fermi energy and carrier density are $E_{F,\mathrm{bulk}}=3240$~meV and $n_0=2.65\times10^{28}$~m$^{-3}$ \cite{Mermin}, yielding $L_c=6.19\,\text{\AA}$.

In this case, despite a modest enhancement of the electron--phonon coupling near $L_c$, the Coulomb repulsion remains dominant. As a result, the calculated maximum critical temperature remains below $5\times10^{-3}$~K, as shown in Fig.~\ref{fig:alkali_tc}, indicating that confinement alone is insufficient to induce an observable superconducting state in sodium.

\subsection*{Potassium}

For potassium, the thickness dependence of $\lambda$ and $\mu^{*}$ is shown in Fig.~\ref{fig:alkali_params} (blue curves). The bulk electron--phonon spectral function, characterized by $\lambda_{\mathrm{bulk}}=0.16$, is shown in the inset of Fig.~\ref{fig:alkali_tc} (blue curve) \cite{a2fPotassium}. The bulk Coulomb pseudopotential is $\mu^{*}_{\mathrm{bulk}}(\omega_c)=0.15$ \cite{muPotassium}, with $\omega_c=28$~meV and $\omega_{\max}=30$~meV. The bulk Fermi energy and carrier density are $E_{F,\mathrm{bulk}}=2120$~meV and $n_0=1.40\times10^{28}$~m$^{-3}$ \cite{Mermin}, giving $L_c=7.66\,\text{\AA}$.

As for sodium, the enhancement of $\lambda$ induced by confinement is insufficient to overcome the Coulomb pseudopotential. Consequently, the calculated critical temperature remains below $5\times10^{-3}$~K across the entire thickness range considered, as shown in Fig.~\ref{fig:alkali_tc}.

\subsection*{Rubidium}

The results for rubidium are displayed in Fig.~\ref{fig:alkali_params} (green curves) and Fig.~\ref{fig:alkali_tc}. The bulk electron--phonon spectral function of Rb, with $\lambda_{\mathrm{bulk}}=0.065$, is shown in the inset of Fig.~\ref{fig:alkali_tc} (green curve) \cite{a2fRubidium}. The bulk Coulomb pseudopotential is $\mu^{*}_{\mathrm{bulk}}(\omega_c)=0.126$ \cite{muPotassium}, with $\omega_c=21$~meV and $\omega_{\max}=22$~meV. The bulk Fermi energy and carrier density are $E_{F,\mathrm{bulk}}=1986$~meV and $n_0=1.15\times10^{28}$~m$^{-3}$ \cite{Mermin}, leading to $L_c=8.18\,\text{\AA}$.

In this case, the electron--phonon coupling remains smaller than the Coulomb pseudopotential for all film thicknesses, as clearly visible in Fig.~\ref{fig:alkali_params}. As a consequence, superconductivity is entirely suppressed and no finite critical temperature is obtained.

\subsection*{Cesium}

Finally, we consider cesium. Figure~\ref{fig:alkali_params} (orange curves) shows the thickness dependence of $\lambda$ and $\mu^{*}$, while the corresponding $\alpha^2F(\Omega)$ is shown in the inset of Fig.~\ref{fig:alkali_tc} (orange curve) \cite{a2fCesium}. The bulk electron--phonon coupling is $\lambda_{\mathrm{bulk}}=0.18$, and the Coulomb pseudopotential is $\mu^{*}_{\mathrm{bulk}}(\omega_c)=0.14$ \cite{muCesium}, with $\omega_c=90$~meV and $\omega_{\max}=92$~meV. The bulk Fermi energy and carrier density are $E_{F,\mathrm{bulk}}=1590$~meV and $n_0=0.91\times10^{28}$~m$^{-3}$ \cite{Mermin}, yielding $L_c=8.84\,\text{\AA}$.

As in the cases of sodium and potassium, confinement-induced renormalization is insufficient to stabilize superconductivity. The calculated critical temperature remains below $2.5\times10^{-2}$~K over the entire thickness range, as shown in Fig.~\ref{fig:alkali_tc}.

\section{Beryllium, magnesium, calcium, strontium, and barium}

With the exception of beryllium, the alkaline-earth metals are not superconducting at ambient pressure. Beryllium itself exhibits superconductivity only at a very low critical temperature ($T_c=0.026$~K). As for the alkali metals, it is therefore of interest to examine whether quantum confinement in ultra-thin films can enhance or induce superconductivity in this class of materials.

\subsection*{Beryllium}

We begin with beryllium. Figure~\ref{fig:alkaline_params} shows the thickness dependence of the electron--phonon coupling constant $\lambda$ and the Coulomb pseudopotential $\mu^{*}$ used in the calculations (black curves). The bulk electron--phonon spectral function $\alpha^2F(\Omega)$ of Be, corresponding to $\lambda_{\mathrm{bulk}}=0.21$, is shown in the inset of Fig.~\ref{fig:alkaline_tc} (black curve) \cite{a2fBe}. The Coulomb pseudopotential is fixed to reproduce the experimental bulk critical temperature, yielding $\mu^{*}_{\mathrm{bulk}}(\omega_c)=0.077$, with a cutoff energy $\omega_c=255$~meV and a maximum phonon energy $\omega_{\max}=260$~meV. The bulk Fermi energy and carrier density are $E_{F,\mathrm{bulk}}=14300$~meV and $n_0=2.47\times10^{28}$~m$^{-3}$, respectively \cite{Mermin}, which determine a critical thickness $L_c=2.94\,\text{\AA}$.

Solving the confinement-modified Eliashberg equations, we find that at $L=L_c$ the critical temperature increases to $T_c\simeq0.10$~K, as shown in Fig.~\ref{fig:alkaline_tc}. Although this enhancement is significant compared to the bulk value, the resulting superconductivity remains confined to a very narrow thickness window.

\subsection*{Magnesium}

The results for magnesium are shown in Fig.~\ref{fig:alkaline_params} (red curves) and Fig.~\ref{fig:alkaline_tc}. The bulk electron--phonon spectral function of Mg, characterized by $\lambda_{\mathrm{bulk}}=0.30$, is shown in the inset of Fig.~\ref{fig:alkaline_tc} (red curve) \cite{a2fMg}. The bulk Coulomb pseudopotential is $\mu^{*}_{\mathrm{bulk}}(\omega_c)=0.16$ \cite{tunnMg}, with $\omega_c=90$~meV and $\omega_{\max}=100$~meV. The bulk Fermi energy and carrier density are $E_{F,\mathrm{bulk}}=7080$~meV and $n_0=8.61\times10^{28}$~m$^{-3}$ \cite{Mermin}, yielding a critical thickness $L_c=4.18\,\text{\AA}$.

As seen in Fig.~\ref{fig:alkaline_params}, the electron--phonon coupling exhibits a modest enhancement in the vicinity of $L_c$. Solving the Eliashberg equations, we find that this enhancement is sufficient to stabilize superconductivity, with the resulting critical temperature shown in Fig.~\ref{fig:alkaline_tc} (red curve). Magnesium therefore represents the most favorable case among the alkaline-earth metals considered, displaying a pronounced confinement-induced superconducting instability.

\subsection*{Calcium}

For calcium, the thickness dependence of $\lambda$ and $\mu^{*}$ is shown in Fig.~\ref{fig:alkaline_params} (blue curves). The bulk electron--phonon spectral function, corresponding to $\lambda_{\mathrm{bulk}}=0.096$, is shown in the inset of Fig.~\ref{fig:alkaline_tc} (blue curve) \cite{a2fCa}. The bulk Coulomb pseudopotential is $\mu^{*}_{\mathrm{bulk}}(\omega_c)=0.167$ \cite{muPotassium}, with $\omega_c=66$~meV and $\omega_{\max}=70$~meV. The bulk Fermi energy and carrier density are $E_{F,\mathrm{bulk}}=4690$~meV and $n_0=4.61\times10^{28}$~m$^{-3}$ \cite{Mermin}, giving $L_c=5.15\,\text{\AA}$.

In this case, the electron--phonon coupling remains smaller than the Coulomb pseudopotential over the entire thickness range considered, as shown in Fig.~\ref{fig:alkaline_params}. As a result, superconductivity is completely suppressed and no finite critical temperature is obtained.

\subsection*{Strontium}

The corresponding results for strontium are reported in Fig.~\ref{fig:alkaline_params} (green curves) and Fig.~\ref{fig:alkaline_tc}. The bulk electron--phonon spectral function of Sr, with $\lambda_{\mathrm{bulk}}=0.19$, is shown in the inset of Fig.~\ref{fig:alkaline_tc} (green curve) \cite{a2fSr}. The bulk Coulomb pseudopotential is $\mu^{*}_{\mathrm{bulk}}(\omega_c)=0.16$ \cite{muCesium}, with $\omega_c=39$~meV and $\omega_{\max}=40$~meV. The bulk Fermi energy and carrier density are $E_{F,\mathrm{bulk}}=3930$~meV and $n_0=3.55\times10^{28}$~m$^{-3}$ \cite{Mermin}, leading to $L_c=5.62\,\text{\AA}$.

As in the calcium case, the Coulomb pseudopotential dominates over the electron--phonon interaction for all thicknesses. Consequently, the calculated critical temperature remains below $5\times10^{-3}$~K, as shown in Fig.~\ref{fig:alkaline_tc}.

\subsection*{Barium}

Finally, we consider barium. Figure~\ref{fig:alkaline_params} (orange curves) shows the thickness dependence of $\lambda$ and $\mu^{*}$, while the corresponding $\alpha^2F(\Omega)$ is shown in the inset of Fig.~\ref{fig:alkaline_tc} (orange curve) \cite{a2fSr}. The bulk electron--phonon coupling is $\lambda_{\mathrm{bulk}}=0.15$, and the Coulomb pseudopotential is $\mu^{*}_{\mathrm{bulk}}(\omega_c)=0.14$ \cite{muCesium}, with $\omega_c=30$~meV and $\omega_{\max}=32$~meV. The bulk Fermi energy and carrier density are $E_{F,\mathrm{bulk}}=3640$~meV and $n_0=3.15\times10^{28}$~m$^{-3}$ \cite{Mermin}, yielding $L_c=5.84\,\text{\AA}$.

As for calcium and strontium, the electron--phonon coupling remains insufficient to overcome Coulomb repulsion. Superconductivity is therefore not stabilized, and the calculated critical temperature remains negligibly small over the entire thickness range.

At the end of this section, Table~\ref{tab:params} summarizes the bulk parameters and the maximum critical temperatures $T_{c,m}$ predicted under confinement for all materials considered.

\section{Comparative outlook}
Taken together, the results for noble metals, alkali metals, and alkaline-earth metals reveal a unified and physically transparent picture of confinement-induced superconductivity in good metals. In all three material classes, quantum confinement reshapes the electronic structure near the Fermi level, leading to a thickness-dependent renormalization of the Fermi energy, the normal density of states, the effective electron--phonon coupling $\lambda$, and the Coulomb pseudopotential $\mu^{*}$. However, the balance between these competing effects is highly material dependent and ultimately determines whether a superconducting instability can be stabilized.

In noble metals (Cu, Ag, Au), confinement induces a modest but crucial enhancement of $\lambda$ near the critical thickness $L_c$, sufficient in some cases to overcome Coulomb repulsion within an extremely narrow thickness window \cite{noblemetal}. Among them, gold stands out as the most favorable case, exhibiting the largest confinement-induced critical temperature, while copper and silver display much lower $T_c$ values and a more severe fine-tuning requirement. This behavior reflects the relatively larger bulk electron--phonon coupling of Au compared to Cu and Ag, combined with comparable Coulomb pseudopotentials.

In alkali metals, despite the presence of confinement-induced renormalization effects similar to those observed in noble metals, superconductivity is largely suppressed. Lithium constitutes a marginal exception, where confinement enhances the critical temperature by more than two orders of magnitude compared to the bulk, yet the absolute value of $T_c$ remains extremely low. For sodium, potassium, rubidium, and cesium, the electron--phonon coupling remains insufficient to overcome the Coulomb pseudopotential over the entire thickness range, preventing the stabilization of a superconducting state.

The alkaline-earth metals exhibit a mixed behavior. Beryllium shows a moderate enhancement of $T_c$ under confinement, while magnesium represents the most favorable case in this class, displaying a pronounced confinement-induced superconducting instability with a sizable critical temperature \cite{Mg}. In contrast, calcium, strontium, and barium remain non-superconducting, as the Coulomb repulsion dominates over the electron--phonon interaction even in the ultra-thin regime. These trends can be traced back to the relative magnitude of the bulk electron--phonon coupling and its sensitivity to confinement-driven renormalization.

Overall, these results demonstrate that quantum confinement alone can, in selected cases, induce superconductivity in elemental metals that are non-superconducting in bulk form. However, this mechanism is intrinsically fine tuned: superconductivity emerges only when a delicate balance between enhanced electron--phonon coupling and Coulomb repulsion is achieved, typically within sub-angstrom thickness intervals around the critical confinement length $L_c$. The strong material dependence and extreme narrowness of the superconducting thickness window highlight the challenges associated with experimentally realizing confinement-induced superconductivity in simple metals, while at the same time providing clear guidelines for identifying the most promising candidate systems.

\section{Combining quantum confinement and proximity effects}

The superconducting proximity effect refers to the modification of the superconducting properties of a material when it is placed in electrical contact with another metal \cite{degennes,Wolf}. The adjacent material may itself be superconducting, normal, or characterized by magnetic or other electronic properties. In its standard formulation, proximity-effect theory typically assumes simple planar geometries and bulk-like electronic structures \cite{McMillan}.

\begin{figure*}[t!]
	\centerline{\includegraphics[width=1\textwidth]{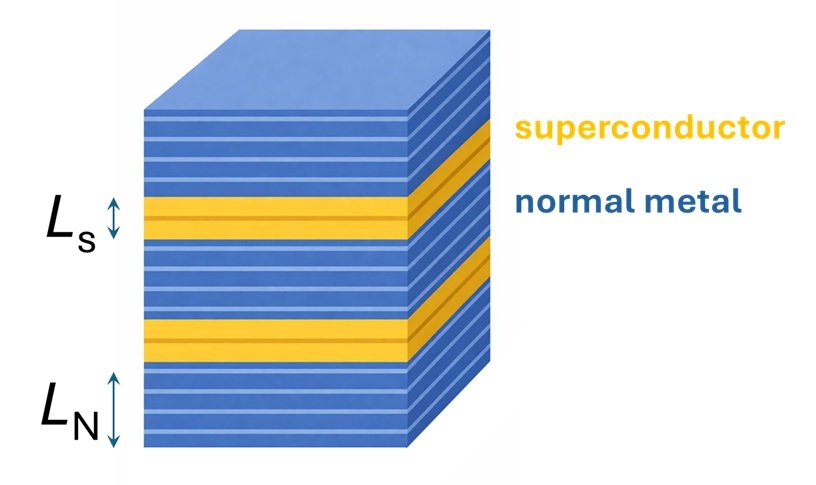}}
\caption{Schematic illustration of a superconductor/normal-metal multilayer system combining quantum confinement and the superconducting proximity effect. 
The superconducting layer ($S$, aluminum in the present case) and the normal-metal layer ($N$, magnesium) have thicknesses $L_S$ and $L_N$, respectively. 
Quantum confinement arises from the finite thickness of each layer, while interlayer coupling enables the proximity-induced transfer of superconducting correlations across the interface.}
\label{fig:bilayer_scheme}
\end{figure*}

Here we extend this framework by combining the superconducting proximity effect with quantum confinement. Rather than considering isolated ultra-thin films—which are experimentally challenging to fabricate at sub-nanometer thickness—we consider layered heterostructures composed of alternating superconducting and normal-metal layers. Such structures can be grown reliably using sputtering or related deposition techniques and allow quantum confinement effects to emerge through the reduced thickness of each constituent layer. A schematic representation of the alternating layers geometry, highlighting the interplay between quantum confinement and proximity coupling, is shown in Fig.~\ref{fig:bilayer_scheme}.

As a representative example, we consider a superconductor/normal-metal bilayer composed of aluminum (Al) and magnesium (Mg). This choice is motivated by two considerations. First, Al and Mg do not form stable alloys under the relevant growth conditions, ensuring that the layers remain chemically distinct, in contrast to combinations such as Au/Ag. Second, both materials are electronically simple, and all parameters required by the theory are well established experimentally or from \textit{ab initio} calculations. For simplicity, we assume equal thicknesses for the superconducting and normal layers, $L_S=L_N$, although the formalism readily allows for unequal thicknesses.

To describe superconductivity in the presence of both proximity coupling and quantum confinement, we solve the isotropic one-band $s$-wave Eliashberg equations generalized to multilayer systems. In this case, four coupled equations must be solved self-consistently for the superconducting gaps $\Delta_{S(N)}(i\omega_n)$ and the mass-renormalization functions $Z_{S(N)}(i\omega_n)$ in the superconducting ($S$) and normal ($N$) layers. The Eliashberg equations on the imaginary axis read \cite{McMillan,Ummaprox1,Ummaprox2,Ummaprox3,Carbi1,Carbi2,Carbi3,Carbi4,kresin}:
\begin{eqnarray}
\omega_{n}Z_{N}(i\omega_{n})=\omega_{n}+ \pi T\sum_{m}\Lambda^{Z}_{N}(i\omega_{n},i\omega_{m})N^{Z}_{N}(i\omega_{m})+\Gamma\ped{N} N^{Z}_{S}(i\omega_{n})
\label{eq:EE1}
\end{eqnarray}
\begin{eqnarray}
&&Z_{N}(i\omega_{n})\Delta_{N}(i\omega_{n})=\pi
T\sum_{m}\big[\Lambda^{\Delta}_{N}(i\omega_{n},i\omega_{m})-\mu^{*}_{N}(\omega_{c})\big]\times\nonumber\\
&&\times\Theta(\omega_{c}-|\omega_{m}|)N^{\Delta}_{N}(i\omega_{m})
+\Gamma\ped{N} N^{\Delta}_{S}(i\omega_{n})
 \label{eq:EE2}
\end{eqnarray}
\begin{eqnarray}
\omega_{n}Z_{S}(i\omega_{n})=\omega_{n}+ \pi T\sum_{m}\Lambda^{Z}_{S}(i\omega_{n},i\omega_{m})N^{Z}_{S}(i\omega_{m})+\Gamma\ped{S} N^{Z}_{N}(i\omega_{n})
\label{eq:EE3}
\end{eqnarray}
\begin{eqnarray}
&&Z_{S}(i\omega_{n})\Delta_{S}(i\omega_{n})=\pi
T\sum_{m}\big[\Lambda^{\Delta}_{S}(i\omega_{n},i\omega_{m})-\mu^{*}_{S}(\omega_{c})\big]\times\nonumber\\
&&\times\Theta(\omega_{c}-|\omega_{m}|)N^{\Delta}_{S}(i\omega_{m})
+\Gamma\ped{S}N^{\Delta}_{N}(i\omega_{n})
 \label{eq:EE4}
\end{eqnarray}
where
\begin{eqnarray}
&&N^{\Delta}_{S(N)}(i\omega_{m})=
\frac{\Delta_{S(N)}(i\omega_{m})}
{\sqrt{\omega^{2}_{m}+\Delta^{2}_{S(N)}(i\omega_{m})}}
\left[\frac{N_{S(N)}(i\omega_{m})+N_{S(N)}(-i\omega_{m})}{2}\right]\times\nonumber\\
&&\times\frac{2}{\pi} \arctan\!\left(\frac{W_{S(N)}}{2Z_{S(N)}(i\omega_{m})\sqrt{\omega_{m}^{2}+\Delta_{S(N)}^{2}(i\omega_{m})}}\right),
\end{eqnarray}
\begin{eqnarray}
&&N^{Z}_{S(N)}(i\omega_{m})=
\frac{\omega_{m}}
{\sqrt{\omega^{2}_{m}+\Delta^{2}_{S(N)}(i\omega_{m})}}
\left[\frac{N_{S(N)}(i\omega_{m})+N_{S(N)}(-i\omega_{m})}{2}\right]\times\nonumber\\
&&\times\frac{2}{\pi} \arctan\!\left(\frac{W_{S(N)}}{2Z_{S(N)}(i\omega_{m})\sqrt{\omega_{m}^{2}+\Delta_{S(N)}^{2}(i\omega_{m})}}\right).
\end{eqnarray}
The pairing kernels are defined as
\begin{equation}
\Lambda_{S(N)}(i\omega_{n},i\omega_{m})=
2\int_{0}^{+\infty}d\Omega\,
\frac{\Omega\,\alpha^{2}_{S(N)}F(\Omega)}
{(\omega_{n}-\omega_{m})^{2}+\Omega^{2}},
\end{equation}
where $\alpha^{2}_{S(N)}F(\Omega)$ are the electron--phonon spectral functions of the superconducting and normal layers. The corresponding electron--phonon coupling constants are
\begin{equation}
\lambda_{S(N)}=
2\int_{0}^{+\infty}d\Omega\,
\frac{\alpha^{2}_{S(N)}F(\Omega)}{\Omega}.
\end{equation}

The coupling between layers is described by
\begin{equation}
\Gamma_{S(N)}=\pi |t|^{2} A L_{N(S)} N_{N(S)}(0),
\label{eq:EE6}
\end{equation}
with $\Gamma_S/\Gamma_N = L_N N_N(0)/(L_S N_S(0))$. Here $A$ is the junction cross-sectional area, $|t|^{2}$ is the transmission probability, $L_{S(N)}$ are the layer thicknesses, and $N_{S(N)}(0)$ are the densities of states at the Fermi level.

As in the single-layer case, quantum confinement induces renormalizations of the key material parameters:
\begin{equation}
\lambda_{S(N)} = C_{S(N)}\,\lambda_{S(N),\mathrm{bulk}},
\end{equation}
\begin{equation}
\mu^{*}_{S(N)}=
\frac{C_{S(N)}\,\mu_{S(N),\mathrm{bulk}}}
{1+\mu_{S(N),\mathrm{bulk}}\ln(E_{F,S(N)}/\omega_c)},
\end{equation}
\begin{equation}
E_{F,S(N)}=C_{S(N)}^{2}E_{F,S(N),\mathrm{bulk}},
\end{equation}
\begin{equation}
N(E_{F,S(N)})=C_{S(N)}\,N_{S(N),\mathrm{bulk}}(0).
\end{equation}

To solve the coupled equations, nine input parameters are required: the two electron--phonon spectral functions $\alpha^{2}_{S(N)}F(\Omega)$, the two Coulomb pseudopotentials $\mu^{*}_{S(N)}$, the two densities of states $N_{S(N)}(0)$, the two layer thicknesses $L_{S(N)}$, and the product $A|t|^{2}$. The geometrical parameters are taken from experiment; we assume $|t|^{2}=1$ (ideal interface) and $A=10^{-7}$~m$^{2}$, noting that the final results are independent of $A$.

For aluminum, we use $\lambda_{S,\mathrm{bulk}}=0.43$ and $\alpha^{2}F(\Omega)$ from Ref.~\cite{nostro}, with $N_{S,\mathrm{bulk}}(0)=0.1757$~eV$^{-1}$ per unit cell. For magnesium, $N_{N,\mathrm{bulk}}(0)=0.2338$~eV$^{-1}$ per unit cell \cite{Butler_N0AlMg}. The Coulomb pseudopotential of aluminum is fixed to reproduce the bulk critical temperature $T_c=1.2$~K, yielding $\mu^{*}_{S,\mathrm{bulk}}=0.13984$ with $\omega_c=145$~meV and $\omega_{\max}=150$~meV. The bulk Fermi energy and carrier density are $E_{F,\mathrm{bulk}}=11700$~meV and $n_{S0}=1.81\times10^{29}$~m$^{-3}$ \cite{Mermin}, giving a critical thickness $L_{cS}=3.26\,\text{\AA}$.

\begin{figure*}[t!]	\centerline{\includegraphics[width=1\textwidth]{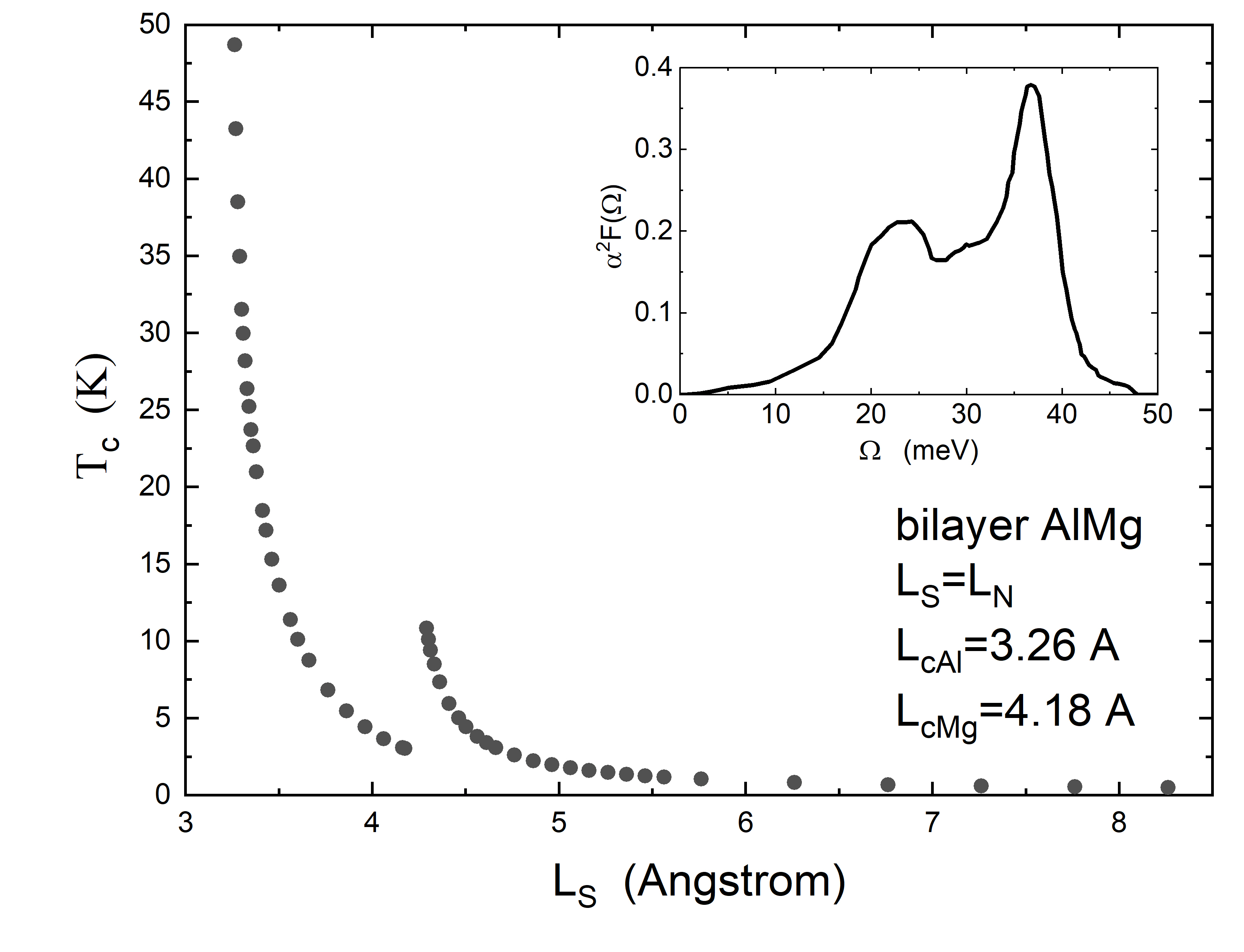}}
\caption{Critical temperature $T_c$ as a function of the single-layer thickness $L_S$ for an Al/Mg superconductor--normal-metal bilayer with equal layer thicknesses ($L_S=L_N$). Full symbols denote numerical solutions of the confinement- and proximity-modified Eliashberg equations. The non-monotonic dependence of $T_c$ reflects the interplay between quantum confinement, which renormalizes the electronic density of states and pairing interaction in each layer, and the superconducting proximity effect across the interface. In selected thickness ranges, confinement-enhanced pairing leads to critical temperatures exceeding the bulk value of aluminum. The inset shows the Eliashberg electron--phonon spectral function $\alpha^2F(\Omega)$ of bulk aluminum used as input in the calculations.}
\label{fig:bilayer_tc}
\end{figure*}

The Eliashberg equations are solved numerically assuming equal layer thicknesses. The resulting critical temperature as a function of thickness is shown in Fig.~\ref{fig:bilayer_tc}. Two key features emerge: first, due to quantum confinement, the critical temperature exceeds the bulk aluminum value in selected thickness ranges; second, the dependence of $T_c$ on thickness is strongly non-monotonic. This behavior arises from the mismatch between the critical confinement lengths $L_{cS}$ and $L_{cN}$, leading to a discontinuity in $T_c(L)$ at $L=L_{cN}$. 

Specifically, Fig.~\ref{fig:bilayer_tc} shows that the critical temperature $T_c$ depends non-monotonically on the layer thickness: at selected thicknesses, quantum confinement independently enhances superconducting pairing in the Mg layer and, simultaneously, strengthens proximity coupling with the Al layer, leading to pronounced maxima in $T_c$. The discontinuity in $T_c$ signals the thickness at which the confinement regimes of the two materials change relative to one another, abruptly modifying the balance between intrinsic confinement-induced superconductivity in Mg and proximity-induced pairing from Al.

These results illustrate how nanoscale heterostructuring can be used to engineer superconducting properties beyond bulk material limitations, by exploiting the cooperative effects of quantum confinement and proximity coupling.


\section{Conclusions and outlook}

In recent years, substantial progress has been made in understanding how quantum confinement can qualitatively reshape superconductivity in metallic systems. In this article, we have reviewed and synthesized these developments within a confinement-generalized Eliashberg framework that enables fully quantitative, parameter-free calculations of superconducting properties in ultra-thin metallic films. The central physical mechanism underlying this approach is the thickness-dependent reconstruction of the electronic structure, which modifies the normal density of states near the Fermi level and induces correlated renormalizations of the Fermi energy $E_F$, the effective electron--phonon coupling, and the Coulomb pseudopotential.

Numerical solutions of the resulting Eliashberg equations demonstrate that superconductivity can, in principle, be induced in selected elemental metals that are non-superconducting in bulk form. However, this instability is intrinsically fine tuned: superconductivity emerges only within extremely narrow thickness intervals, typically centered around sub-nanometer length scales, where the enhancement of the electron--phonon interaction is sufficient to overcome Coulomb repulsion. This sensitivity highlights both the experimental challenges associated with realizing confinement-induced superconductivity in good metals and the importance of precise thickness control. At the same time, it provides a clear diagnostic criterion for disentangling genuine confinement-driven effects from superconductivity arising from proximity, disorder, or interface-induced mechanisms.

Beyond isolated films, we have shown that combining quantum confinement with the superconducting proximity effect in layered superconductor/normal-metal heterostructures offers a more robust and experimentally accessible route to enhancing superconductivity. In such systems, confinement and interlayer coupling act cooperatively, leading in selected thickness ranges to critical temperatures that exceed those of the bulk superconducting constituent. The resulting non-monotonic dependence of $T_c$ on layer thickness reflects the interplay between distinct confinement length scales in the superconducting and normal layers, and points to new strategies for engineering superconducting properties through nanoscale design.

Advances in high-pressure techniques have uncovered superconductivity at record critical temperatures in compressed hydrides and elemental systems, notably in sulfur hydride where \(T_c\sim203\) K was observed under megabar pressures \cite{DrozdovHemley2015}.

A particularly intriguing perspective emerges when comparing confinement-induced superconductivity with pressure-induced superconductivity in elemental metals. 
Recent experiments have demonstrated that elemental scandium reaches a record-high critical temperature of approximately $36$~K under extreme compression ($P \sim 260$~GPa), driven by pressure-induced electronic reconstruction and strong electron--phonon coupling in high-pressure phases \cite{ScandiumPRL}. 
From a theoretical standpoint, high pressure modifies superconductivity primarily by increasing orbital overlap, inducing $s$--$d$ charge transfer, reshaping the Fermi surface, and renormalizing phonon spectra, effects that can be quantitatively captured within first-principles \cite{Monacelli} and Eliashberg-based approaches \cite{SettyPRB103,Margine}.
Within this context, the confinement framework discussed in the present work may be viewed as complementary to pressure-based tuning: both mechanisms act by reshaping the electronic phase space near the Fermi level and enhancing the effective pairing interaction. 
This suggests that a unified description, combining confinement-induced density-of-states engineering with theoretical treatments of pressure-driven phonon and electronic renormalization, could provide a powerful route for rationally designing superconductivity in elemental and low-complexity metallic systems.

Looking ahead, several natural extensions of this framework are anticipated. These include generalizations to multiband superconductors, where confinement-induced shape resonances and interband coupling may play a decisive role, to unconventional pairing symmetries such as $d$-wave order parameters, and to the inclusion of disorder and interface scattering, which are unavoidable in realistic thin-film and heterostructured systems. Together, these directions define a broad and fertile research landscape in which quantum confinement, electronic structure, and many-body interactions intersect, offering new opportunities for the controlled design and discovery of superconductivity at the nanoscale.

Finally, the ability to manipulate superconducting properties via external fields has direct implications for future superconducting nano-electronics and quantum technologies. Recent work has shown that the supercurrent in metallic thin films and Josephson nanotransistors can be modulated or even suppressed by an applied DC electric field \cite{fomin}, forming the basis of supercurrent field-effect transistors \cite{giazo,Zaccone2025_PRB111,UmmarinoPRMaterialsSupercurrent} and field-tunable superconducting elements with potential for low-power digital and cryogenic circuits \cite{forn,giazo}. Such devices could serve as building blocks for superconducting logic and memory, where gate-controlled critical currents enable fast switching. Moreover, superconducting qubits — including transmons and flux qubits based on Josephson junctions — exploit macroscopic quantum coherence of supercurrents for quantum information processing, and are a leading platform for scalable quantum computing \cite{Krantz2019_QuBits}. The interplay between quantum confinement, field-effect control, and qubit architectures therefore opens new avenues for integrating material-level engineering with quantum device functionality.\\

\noindent\textbf{Data availability.} All data that support the findings of this study are included within the article.

\section*{Acknowledgements}
We thank prof G. Ubertalli for useful discussions.
\newpage

\end{document}